# Rapid analysis of point-contact Andreev reflection spectra via machine learning with adaptive data augmentation

Dongik Lee[1], Valentin Stanev[2], Xiaohang Zhang[2], Mijeong Kang[3], Ichiro Takeuchi[2,4], and Seunghun Lee[1*]

[1]Department of Physics, Pukyong National University, Busan 48513, Republic of Korea

[2]Department of Materials Science and Engineering, University of Maryland, College Park, Maryland 20742, United States

[3]Department of Optics and Mechatronics Engineering, Pusan National University, Busan 46742, Republic of Korea

[4]Maryland Quantum Materials Center, Department of Physics, University of Maryland, College Park, Maryland 20742, United States

*Corresponding author E-mail: seunghun@pknu.ac.kr (Seunghun Lee)

## Abstract

Delineating the superconducting order parameters is a pivotal task in investigating superconductivity for probing pairing mechanisms, as well as their symmetry and topology. Point-contact Andreev reflection (PCAR) measurement is a simple yet powerful tool for identifying the order parameters. The PCAR spectra exhibit significant variations depending on the type of the order parameter in a superconductor, including its magnitude ($\Delta$), as well as temperature, interfacial quality, Fermi velocity mismatch, and other factors. The information on the order parameter can be obtained by finding the combination of these parameters, generating a theoretical spectrum that fits a measured experimental spectrum. However, due to the complexity of the spectra and the high dimensionality of parameters, extracting the fitting parameters is often time-consuming and labor-intensive. In this study, we employ a convolutional neural network (CNN) algorithm to create models for rapid and automated analysis of PCAR spectra of various superconductors with different pairing symmetries (conventional $s$-wave, chiral $p_x+ip_y$-wave, and $d_{x^2-y^2}$-wave). The training datasets are generated based on the Blonder-Tinkham-Klapwijk (BTK) theory and further modified and augmented by selectively incorporating noise and peaks according to the bias voltages. This approach not only replicates the experimental spectra but also brings the model's attention to important features within the spectra. The optimized models provide fitting parameters for experimentally measured spectra in less than 100 ms per spectrum. Our approaches and findings pave the way for rapid and automated spectral analysis which will help accelerate research on superconductors with complex order parameters.

## Introduction

Superconductivity, characterized by the zero resistance and the Meissner effect, continues to attract significant attention in condensed matter physics. The development of the Bardeen-Cooper-Schrieffer (BCS) theory provided a microscopic understanding of conventional superconductors (SCs)[1]. However, the high-$T_c$ SCs, such as cuprate[2] and iron-based SCs[3], as well as topological SCs – which are candidates for generating fault-tolerant qubits – cannot be described by the BCS theory[4]. Recently, numerous SCs have been theoretically predicted and experimentally confirmed[5-7].

The SCs are classified as either conventional or unconventional SCs based on the formation mechanisms of Cooper pairs and the symmetry of their order parameters. Most conventional SCs, explained by BCS theory based on electron-phonon interaction, have an *s*-wave order parameter. In contrast, unconventional SCs, which BCS theory or its extension (*e.g.*, Eliashberg theory) do not account for, exhibit different order parameters. For instance, the cuprate SCs have $d_{x^2-y^2}$-wave order parameters, and chiral *p*-wave, one of the topological SCs, have $p_x+ip_y$-wave order parameters.

To investigate the properties and order parameters of SCs, various experimental techniques have been employed, including angle-resolved photoemission spectroscopy (APRES)[8], scanning tunneling microscopy (STM)[9], nuclear magnetic resonance (NMR)[10], and point-contact Andreev reflection (PCAR) spectroscopy[11], as well as systematic temperature and/or field dependent measurements of resistance[12], magnetization[13], and specific heat[14]. Among these, PCAR spectroscopy is a convenient and powerful tool to probe spectroscopic information (especially superconducting gap) of SCs[15-20], and it does not require large size samples or atomically flat surfaces[21]. The PCAR spectroscopy is based on the phenomenon of Andreev reflection where an electron incident on a normal metal-superconductor interface results in a hole with opposite spin and velocity reflecting to conserve the spin, momentum and charge, since electrons and Cooper pairs are current carriers in the normal metal and the superconductor, respectively. When electrons flow through the point-contact ballistically (*i.e.*, without scattering), ideally, Andreev reflection doubles the conductance when the energy is less than the superconducting gap energy ($\Delta$). In practical experiments, however, nonideal contacts in a thermal regime, a barrier potential, and the Fermi velocity mismatch ($Z$) between the metal and superconductor have influences on measured spectra. The Blonder-Thinkham-Klapwijk (BTK) theory incorporates these parameters to describe the differential conductance as a function of bias (*i.e.*, PCAR spectra)[22]. For more practical analysis, broadening parameter ($\Gamma$) in PCAR spectra caused by the finite lifetime of quasiparticles should be considered[23]. The PCAR spectra of various shapes can be constructed by adjusting such parameters including the temperature ($T$), $\Delta$, $Z$, and $\Gamma$ parameters (Fig. S1). In addition, electron indent direction also needs to be considered for SCs having spatially anisotropic order parameters such as $p_x+ip_y$- and $d_{x^2-y^2}$-wave[24-26].

By fitting experimental PCAR spectra with theoretical spectra simulated by the BTK theory, we can extract information about the superconducting order parameter. However, the fitting process is often time-consuming and labor-intensive. The presence of inevitable noise – such as thermal and shot noise – and dips caused by nonideal contacts (such as those in the thermal regime and diffusive regime) complicates the fitting process further[27,28]. To ensure reproducible and reliable results, multiple samples and measurements are often required, which inevitably extend the time needed for analysis[7,29,30].

In this study, we demonstrate a straightforward and rapid method for analyzing PCAR spectra using convolutional neural network (CNN) machine learning (ML) models. While the potential for applying ML to various materials characterization data, such as x-ray diffraction patterns[31-33], x-ray photoelectron spectra[34], pair distribution functions[35], Raman spectra[36], and ellipsometry spectra[37], have been suggested, it has not been previously utilized for the analysis of PCAR spectra. We train the CNN models using theoretical spectra generated through (extended and modified) BTK models for three different order parameters ($s$-, $p_x+ip_y$-, and $d_{x^2-y^2}$-wave) with various parameter combinations, as illustrated in Fig. 1. The models are then applied to analyze experimental spectra to identify the best fitting parameters. This study represents the first demonstration of ML applied to the analysis of PCAR spectra. Additionally, our study provides guidelines for data preprocessing and augmentation for practical ML-based spectral analysis in general.

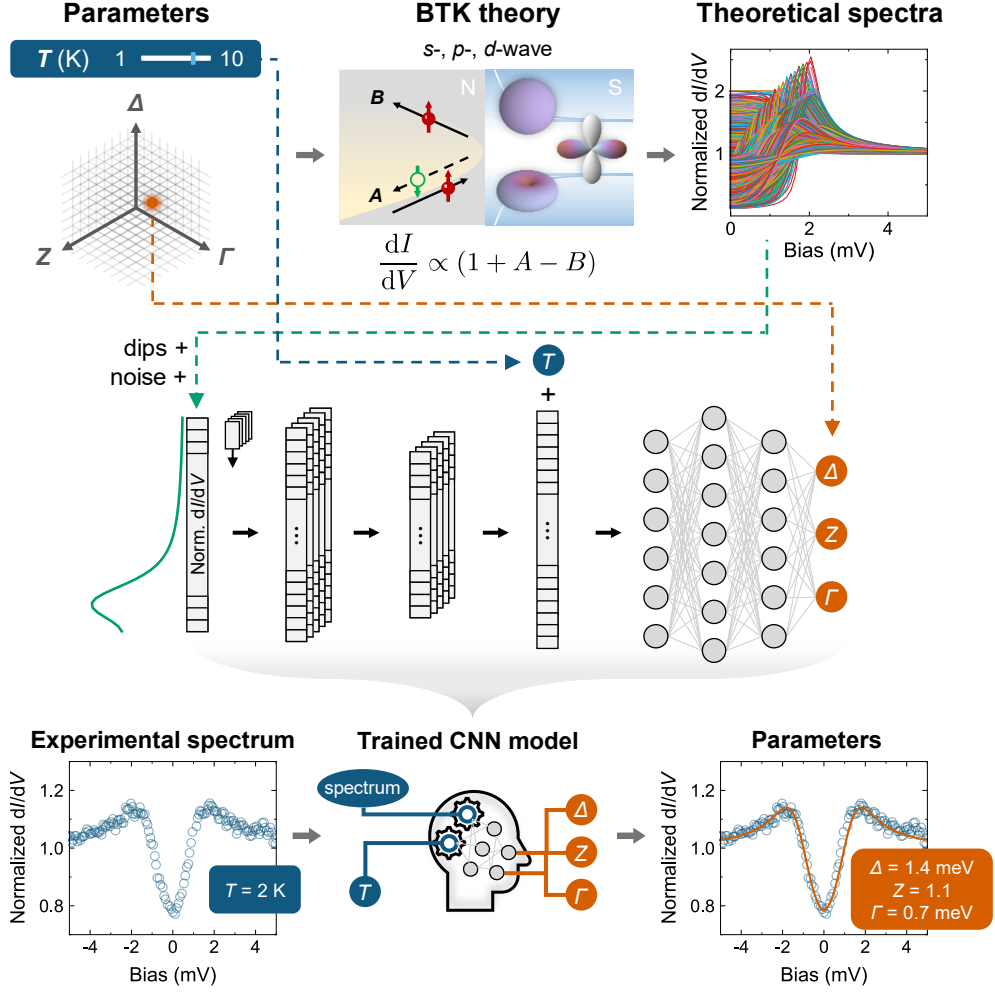

**Fig. 1 | Workflow for analyzing point-contact Andreev reflection (PCAR) spectra using machine learning.** The workflow consists of simulating PCAR spectra (top), training the model (middle), and analyzing experimental data (bottom). Theoretical spectra are simulated under various parameters using extended and modified Blonder-Tinkham-Klapwijk (BTK) theory for the interfaces between $s$-, $p_x+ip_y$-, and $d_{x^2-y^2}$-wave superconductors (S) and a normal metal (N). Here, A and B denote the probabilities of hole reflection (Andreev reflection) and electron reflection, respectively. Temperature and simulated spectra are set as the input data for the convolutional neural network models, while other parameters are set as the output data. The trained models are used to analyze experimental data and obtain the best fitting parameters.

# Results

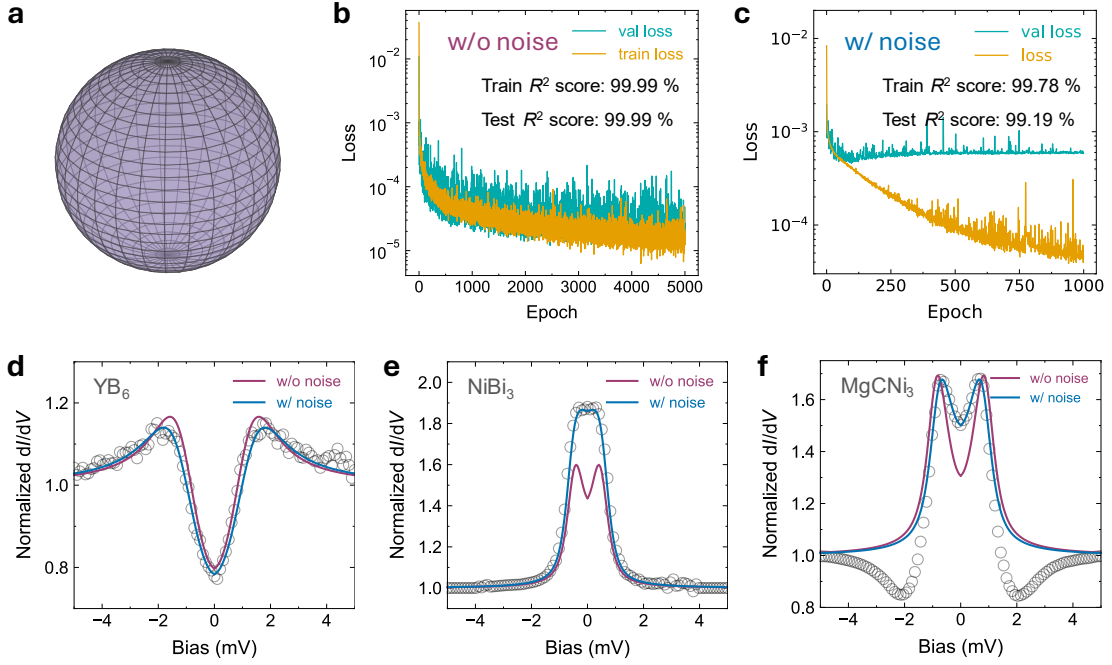

**Fig. 2 | Analysis of point-contact Andreev reflection (PCAR) spectra for the *s*-wave superconductors using convolutional neural network (CNN) models.** (a) Schematic representation of *s*-wave pair potential in k-space. The *s*-wave pair potential is isotropic. (b) and (c) are learning curves of CNN models trained without noise (w/o noise) and with noise and dips (w/ noise), respectively. The orange lines and green lines represent training loss and validation loss, respectively. (d)–(f) Predictions using w/o noise (yellow lines) and w/ noise (blue lines) models for experimental PCAR spectra (open circles): $YB_6$[7], $NiBi_3$[38], and $MgCNi_3$[39], respectively. The predicted parameters used for simulating theoretical spectra w/ noise (blue lines) are $T$ = 2 K, $\Delta$ = 1.37 meV, $Z$ = 1.07, and $\Gamma$ = 0.692 meV for $YB_6$, $T$ = 1.51 K, $\Delta$ = 0.649 meV, $Z$ = 0.166, and $\Gamma$ = 0.000 meV for $NiBi_3$, $T$ = 1.7 K, $\Delta$ = 0.911 meV, $Z$ = 0.339, and $\Gamma$ = 0.020 meV for $MgCNi_3$. The other fitting parameters (from the w/o noise model and the references) are summarized in Table S2.

*Analysis of PCAR spectra for conventional s-wave superconductors*

We developed ML models to analyze PCAR for *s*-wave SCs with an isotropic order parameter (Fig. 2a). To generate the training data for the models, we simulated PCAR spectra under various conditions of $T$, $\Delta$, $Z$, and $\Gamma$ using BTK theory. The simulation parameters are summarized in Table S1.

We compared CNN models trained using theoretically simulated spectra as is (w/o noise model) and modified theoretical spectra (w/ noise model) by adding experimental features, which are usually ignored in the fitting process. Here, the experimental features include white noise (such as thermal and shot noise) and conductance dips caused by the finite resistivity of the SCs (*i.e.*, Maxwell term) for current passing through some of the contacts that are not in the ballistic limit (*i.e.*, thermal regime and diffusive regime) above the superconducting critical current (see Method section for details)[27,28]. We divided the dataset (originally 47,187 spectra for *s*-wave) into training and test sets in an 8:2 ratio. The training set was used for model training using a 5-fold cross-validation technique, and we evaluated both training and test scores. The optimal hyperparameters were determined by tuning various parameters (such as the number of convolutional and dense layers, epochs, and the batch size) to maximize the performance score. The w/o noise model demonstrates nearly perfect predictions, achieving training and test $R^2$ scores of 99.998% and 99.985%, respectively (Fig. 2b). Such the high scores indicate that the CNN model here are able to capture the essential features of the BTK model nearly perfectly. In comparison, the w/ noise model exhibits slightly reduced training and test $R^2$ scores of 99.8% and 99.2%, respectively, but still show well-trained performance exceeding 99% (Fig. 2c). Adding the experimental features – the data preprocessing and augmentation increase the complexity of the data, which leads to such the lower performance scores, but we expected that these processes enhance robustness against experimental artifacts not reproduced by BTK simulation and thus improve the practical applicability and generality of the CNN model for real experimental data[40]. To validate the performance of the models, we applied them to experimental PCAR spectra of various *s*-wave SCs obtained from the literature and evaluated the predicted parameters (Table S2)[7,38,39]. We assessed the models' performance by comparing experimental spectra with theoretical spectra simulated using the predicted parameters.

Figures 2d–f show the comparison of experimental spectra and theoretical spectra simulated using the predicted parameters with the w/o noise and w/ noise models. The experimental spectra in Figs. 2d and 2e were selected as representative spectra measured under two different $Z$ regimes – tunneling ($Z > 1$) and intermediate ($Z \simeq 1$) regimes in which the differential conductance (d$I$/d$V$) near zero bias is less or greater than 1, respectively. The experimental spectrum in Fig. 2f includes conductance dips, possibly due to the nonideal contacts as discussed above. It is evident that while both models predict the similar gap values, the w/ model provides a much more accurate fit. Notably, artifacts such as dips in the experimental data complicate using quantitative error metrics such as the mean absolute error (MAE) for the fitting process (see Fig. S2 for details). This further underscores the utility of using ML for PCAR analysis.

The theoretical spectra simulated using parameters predicted by the w/o noise model often fail to match the experimental spectra near the zero-bias conductance region. In contrast, the w/ noise model successfully reproduces the experimental features, indicating that training with the adaptively augmented dataset – comprising modified theoretical spectra that include noise and artifacts absent in standard BTK theory – improves the model's practical performance in ways not directly reflected by $R^2$ scores. We hypothesize that training the model with the adaptively augmented dataset provides two key benefits: (1) it makes the training data more realistic, thereby reducing the influence of noise and artifacts present in experimental spectra (enhancing robustness), and (2) it effectively increases the size of the training dataset through data augmentation[41].

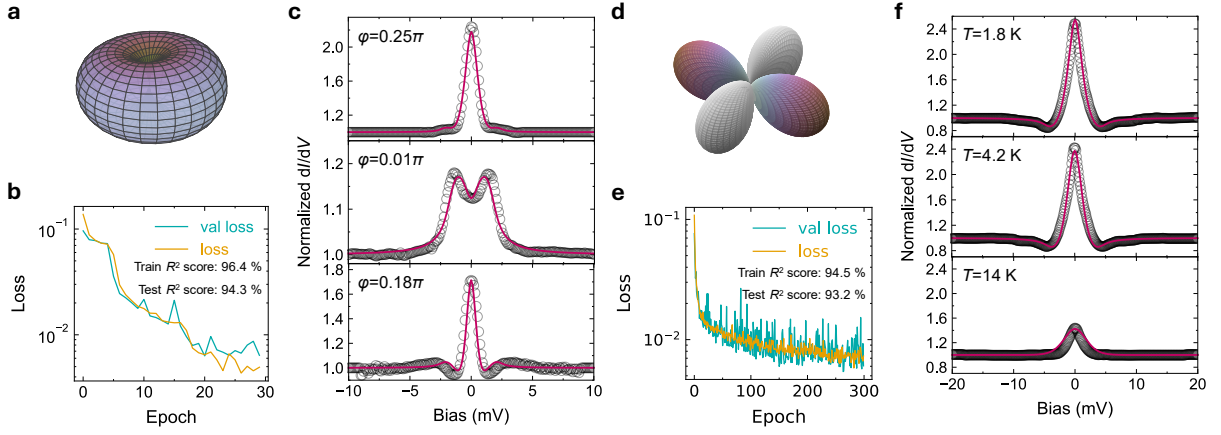

**Fig. 3 | Analysis of PCAR spectra for unconventional superconductors using CNN models.** (a) Schematic representation of $p_x+ip_y$ (*i.e.*, chiral *p*-wave) pairing potential. (b) Learning curves of CNN models trained with noise. The orange lines and green lines represent training loss and validation loss, respectively. (c) PCAR spectra of Bi/Ni bilayer measured in different point-contact directions[42]. The top-left parameters indicate the angle between the $k_x$-axis of the order parameter and the normal direction of the interface (*i.e.*, electron incidence direction) in the PCAR measurements. The open circles and red lines represent experimental PCAR spectra and theoretical PCAR spectra predicted by the CNN model, respectively. The predicted parameters used for simulating theoretical spectra (red curves) are $T = 1.43$ K, $\Delta = 2.45$ meV, $Z = 0.861$, $\Gamma = 0.132$ meV, and $\varphi = 0.25\pi$ (top); $T = 1.43$ K, $\Delta = 1.98$ meV, $Z = 0.344$, $\Gamma = 0.189$ meV and $\varphi = 0.01\pi$ (middle); $T = 1.43$ K, $\Delta = 2.33$ meV, $Z = 1.22$, $\Gamma = 0.212$ meV, and $\varphi = 0.18\pi$ (bottom). (d) Schematic representation of $d_{x^2-y^2}$-wave pairing potential. (e) Learning curves of CNN models trained with noise. The orange lines and green lines represent training loss and validation loss, respectively. (f) PCAR spectra of PuCoGa$_5$ measured at different temperatures (1.8, 4.2, and 14 K, top to bottom)[24]. Open circles and red lines represent experimental PCAR spectra and theoretical PCAR spectra predicted by the CNN model, respectively. The top-left parameters indicate the temperatures at which each PCAR spectrum was measured. The predicted parameters for these curves are $T = 1.8$ K, $\Delta = 4.76$ meV, $Z = 0.799$, $\Gamma = 0.478$ meV, and $\varphi = \pi/4$ (top); $T = 4.2$ K, $\Delta = 4.23$ meV, $Z = 0.893$, $\Gamma = 0.430$ meV, and $\varphi = \pi/4$ (middle); $T = 14$K, $\Delta = 2.11$ meV, $Z = 0.599$, $\Gamma = 0.000$ meV, and $\varphi = \pi/4$ (bottom).

*Analysis of PCAR spectra for unconventional superconductors*

We built a model to analyze PCAR spectra of the $p_x+ip_y$ (chiral *p*-wave) triplet SCs. The chiral *p*-wave pairing symmetry (Fig. 3a) is one of the candidates for topological superconductivity[43-45], which is capable of hosting Majorana fermions and holds promise for quantum computing applications[4,46-48]. Therefore, identifying the chiral *p*-wave has been a pivotal yet very challenging issue. The chiral *p*-wave order parameter consists of two *p*-wave components: a real term and an imaginary term of equal magnitudes. The imaginary term breaks time-reversal symmetry (**T**:$p_x+ip_y \mapsto p_x-ip_y$) and exhibits fully gapped superconductivity with a phase that varies depending on the direction. The angle-dependent pair potential indicates that the order parameter is a function of the angle ($\varphi$), defined as the angle between the $k_x$-axis of the order parameter and the normal direction of the interface (*i.e.*, electron incidence direction). The PCAR spectra of chiral *p*-wave SCs are simulated using pairing potentials of $\Delta\sin(\varphi)$. Consequently, the PCAR spectra for chiral *p*-wave SCs exhibit more complex conductance curves compared to those of *s*-wave SCs (see Fig. S3 in Supplementary Information).

We extended the methodology applied to build the model for *s*-wave SCs to unconventional SCs. We utilized a CNN architecture similar to that used for *s*-wave cases, modifying hyperparameters such as the number of convolution layers and epochs to optimize the models for unconventional SCs. The simulation parameters for generating the training and test datasets for each pairing symmetry are summarized in Table S1. In PCAR spectra of unconventional SCs, conductance dips, which are similar to the feature possibly due to the nonideal contact, can emerge, as shown in the theoretical spectra simulated by modified BTK theory (black arrows in Fig. S3). Therefore, arbitrarily introducing artifact dips features can lead to missing critical features in model predictions. For this reason, we trained each model using an extended dataset that includes only white noise whose amplitude was selectively applied according to the bias voltages: ±0.001 for the inside and ±0.01 for the outside of the $\Delta$ value. This approach helps to avoid the distortion of the zero-bias features in theoretical spectra and guides the model toward accurate and reliable predictions for capturing these features in experimental spectra (see Fig. S4 for details).

The training and test scores of the optimized model for the chiral *p*-wave trained with the augmented dataset are found to be 96.4 % and 94.3 %, respectively (Fig. 3b). The relatively low score compared to the model for *s*-wave is due to the inherent complexity of PCAR spectra for the chiral *p*-wave SCs. To evaluate the model performance, we applied it to experimental PCAR spectra of the Bi/Ni bilayer, which has been suggested to be a chiral *p*-wave SC[42]. Figure 3c demonstrates that predictions are accurate for the spectra obtained from different $\varphi$ measurement angles. Notably, at the bottom of Fig. 3c ($\varphi = 0.18\pi$), the theoretical spectrum from the prediction captures the experimental conductance values inside the $\Delta$ rather than the outside. This is attributed to the selective noise addition in the data preprocessing. The training model with this augmented dataset with selective data preprocessing allows the model to focus on regions with relatively small variations. These findings suggest a strategic approach to data generation and augmentation, directing the model's attention to significant and interesting regions.

We also investigated the applicability of our approach to high-$T_c$ SCs such as cuprates and iron-based SCs with a $d_{x^2-y^2}$ order parameter. The $d_{x^2-y^2}$ order parameter features four lobs and nodes (Fig. 3d). Accordingly, the PCAR spectra are simulated using pairing potentials of $\Delta\cos(2\varphi)$. Following the methodology similar to that for $p_x+ip_y$-wave SCs, but with modified hyperparameters, we built a model for $d_{x^2-y^2}$-wave SCs. This model achieves the training and test scores of 94.5 % and 93.2 %, respectively (Fig. 3e). We then applied the model to analyze the experimental PCAR spectra of $d_{x^2-y^2}$-wave SC PuCoGa$_5$, which was measured at different temperatures with a fixed angle of incidence ($\varphi = \pi/4$)[24]. The predicted parameters obtained from the optimized model reproduce the temperature-dependent PCAR spectra (Fig. 3f), consistent with the values in the literature.

## Discussion

PCAR spectroscopy is a well-known technique for analyzing the order parameters of SCs, and we can extract the information by fitting the spectra using BTK theory. Typically, one takes a trial-and-error method – manually adjusting the fitting parameters to find a theoretical spectrum that (apparently) reproduces the experimental spectrum. While this method works well, it is time-consuming and requires constant human supervision. Using our computer setup (CPU: Intel Ultra 7 155H, GPU: UVIDIA GeForce RTX 4060), the simulation of a single spectrum takes approximately 30 seconds for an $s$-wave SC, but approximately 10 minutes for $p_x+ip_y$-wave and $d_{x^2-y^2}$-wave SCs since the double integration is required for them. The extent of manual intervention may vary depending on the individual's level of expertise and intuition. An alternative approach can be (automated) precomputed spectrum matching based on error metrics such as root mean square error or MAE. Although it may take a long time to construct a database, the fitting process itself can be very rapid. However, the accuracy of this approach is inherently dependent on the database, meaning that insufficient or biased precomputed spectra may lead to incorrect interpretations. Furthermore, if the experimental spectrum contains errors or artifacts that are not present in the ideal theoretical spectra (noise and conductance dips, as discussed above), the error-metric-based analysis may become less reliable or even misleading (Fig. S2).

On the other hand, our models provide very accurate predictions in less than 100 ms per spectrum. To build such a model, our data preprocessing and augmentation play a critical role. By selectively adding additional features to specific regions of the training data, we can control how much attention the model pays to those regions. For instance, if we intend to reduce the model's focus on a specific voltage-bias region, we generate additional training data by introducing noise and distortion to that region while keeping the output (target) unchanged. This approach differs from conventional data augmentation. As demonstrated, we successfully developed a model capable of analyzing experimental spectra, including measurement errors and artifacts (Fig. 2f). By incorporating noise whose level is selectively applied based on the bias voltage, we can guide the model's focus toward d$I$/d$V$ features within $|V| < \Delta$, thereby leading to predictions that accurately reproduce d$I$/d$V$ curves in this region (Figs. 3 and S4). This prediction model mimics human data fitting by emphasizing important regions while occasionally disregarding errors or artifacts. This flexibility highlights the necessity of applying ML for the efficient and accurate analysis of scientific data in general. Indeed, our findings provide guidelines for maximizing the learning efficiency to improve the prediction performance of models. Importantly, our strategy is not limited to PCAR analysis; it can be applied to any type of spectral analysis.

We also anticipate that ML-based PCAR spectral analysis will enable the rapid characterization of new SCs with unknown order parameters, thereby accelerating research in superconductivity. Additionally, since the analysis can be executed without human supervision, this approach has the potential to be implemented as an automated and autonomous system. The current study focused on PCAR spectra of single order parameters ($s$-, $p_x+ip_y$-, or $d_{x^2-y^2}$-wave). However, we expect that the proposed methodology can be extended to analyze the PCAR spectra of multi-gap SCs with mixed order parameters such as MgB$_2$[49]. Moreover, PCAR spectroscopy is widely used as a tool for analyzing the spin polarization of materials[18,50-53]. By incorporating parameters related to spin polarization into the theoretical spectra, PCAR measurements and ML-based spectral analysis demonstrated here can also be used to explore magnetic materials for spintronics.

## Methods

*PCAR spectra simulation for dataset preparation*

The Blonder-Tinkham-Klapwijk (BTK) theory, based on the Bardeen-Cooper-Schrieffer (BCS) theory and the Bogoliubov-de Gennes formalism, describes the bias-dependent differential conductance that arises from Andreev reflection at the interface between a normal metal and a superconductor with an isotropic order parameter (*i.e.*, *s*-wave symmetry)[22]. In SCs with anisotropic order parameters, electrons experience different pair potentials depending on their incident direction (*i.e.*, k-dependency). We let $\varphi$ be the angle between the $k_x$-axis of order parameters and the normal direction of the interface. While $\Delta$ remains constant in *s*-wave SCs, it varies as $\Delta\sin(\varphi)$ and $\Delta\cos(2\varphi)$ in $p_x+ip_y$-wave and $d_{x^2-y^2}$-wave, respectively. Additionally, the BTK theory assumes that current injections occur only perpendicular to the interface (*i.e.*, 1D BTK). However, in reality, charge carriers have an angular distribution. The 1D BTK model provides reliable parameter estimation for *s*-wave SCs, but it is insufficient for unconventional SCs due to their angle-dependent pair potential. Therefore, in such cases, the incident angle of the electron ($\theta_N$), which is the angle between the direction of the incident electron and the normal to the interface, was considered for valid differential conductance[49].

In PCAR spectra, a smaller bias step size allows for better resolution of peaks and fine spectral features, but it also increases the simulation time (Fig. S5). Since the $\Delta$ was set with 0.1 meV intervals in the simulation parameters for data preparation, the step size was set to 0.1 mV to clearly distinguish the spectral change due to the $\Delta$ difference. The bias voltage range was set to 5, 10, and 20 mV, which were found to be typical bias voltage ranges of the PCAR measurement for known *s*-, $p_x+ip_y$-, and $d_{x^2-y^2}$-wave SCs, respectively.

*Data preprocessing*

The experimental spectra may contain experimentally inevitable errors and artifacts, such as noise from measurement equipment and dips from nonideal contacts. To achieve reasonable spectral analysis, we intentionally manipulated the theoretical data to make it more realistic, thus excluding these artifacts in parameter prediction. We introduced the white noise ranging from −0.02 – 0.02 and the dips for the simulated *s*-wave spectra (Fig. S6a). For each theoretical spectrum, we generated five spectra with randomly added noise to ensure that these noise features were disregarded during prediction. Subsequently, we incorporated artificially generated dips by subtracting two symmetric Gaussian functions (Fig. S6b). We created five additional spectra by varying the amplitude of the Gaussian function per each theoretical spectrum.

For unconventional SCs (*i.e.*, $p_x+ip_y$-, and $d_{x^2-y^2}$-wave in this study), the artificial dips were not introduced since conductance dips are theoretically expected from extended BTK theory for unconventional SCs (Fig. S3); instead, only the white noise was incorporated and its level were selectively adjusted according to the bias voltages. We added the white noise to theoretical spectra: ±0.001 for $|V| < \Delta$ and ±0.01 for $|V| > \Delta$. We generated ten spectra with varying noise for each original theoretical spectrum, which were then used as training data.

*CNN architecture*

We implemented the CNN algorithm to predict parameters of PCAR spectra (*i.e.*, $\Delta$, $Z$, and $\Gamma$) from input $T$ and spectra (Fig. 1). In unconventional SCs, different pairing potentials arise depending on the direction, requiring prior knowledge of the crystal orientation and point-contact geometry for accurate PCAR analysis[30]. Therefore, $\varphi$ was included as an additional input parameter. For spectral analysis using one-dimensional matrices, we implemented a 1D CNN model using TensorFlow and Keras libraries in Python. The CNN models consist of two (for $p_x+ip_y$) or four (for *s*- and $d_{x^2-y^2}$-wave) convolutional layers with same padding and ReLU activation functions. We determined the appropriate number of convolutional layers by increasing the layers from zero (*i.e.*, neural network) and evaluating the performance of the model in predicting experimental PCAR spectra. Max pooling layers with size 2 were applied after every two convolutional layers. The resulting feature maps were flattened and concatenated with the temperature (reshaped as a 1×1 matrix) before being passed to the fully connected layers. The fully connected network consists of less than ten layers, each employing ReLU activation functions. To predict the three target parameters (*i.e.*, $\Delta$, $Z$, and $\Gamma$), the output layer was set as three nodes with linear activation functions. The model was trained using the adaptive moment estimation (ADAM) optimizer, with mean squared error as the loss function and a learning rate of 0.001. The batch size was set to 32 (*s*-wave) and 50 ($p_x+ip_y$- and $d_{x^2-y^2}$-wave).